\documentclass[twocolumn, amsmath, amssymb, showkeys]{revtex4-1}

\usepackage[hidelinks]{hyperref}
\usepackage{graphicx}
\usepackage{bbm}
\usepackage{natbib}

\begin{document}

\title{An analysis of the evolution of science-technology linkage in biomedicine}

\author{Qing Ke}
\email{q.ke@northeastern.edu}
\affiliation{Northeastern University, Boston, MA 02115, USA}

\begin{abstract}
Demonstrating the practical value of public research has been an important subject in science policy. Here we present a detailed study on the evolution of the citation linkage between life science related patents and biomedical research over a 37-year period. Our analysis relies on a newly-created dataset that systematically links millions of non-patent references to biomedical papers. We find a large disparity in the volume of science linkage among technology sectors, with biotechnology and drug patents dominating it. The linkage has been growing exponentially over a long period of time, doubling every 2.9 years. The U.S. has been the largest producer of cited science for years, receiving nearly half of the citations. More than half of citations goes to universities. We use a new paper-level indicator to quantify to what extent a paper is basic research or clinical medicine. We find that the cited papers are likely to be basic research, yet a significant portion of papers cited in patents that are related to FDA-approved drugs are clinical research. The U.S. National Institute of Health continues to be an important funder of cited science. For the majority of companies, more than half of citations in their patents are authored by public research. Taken together, these results indicate a continuous linkage of public science to private sector inventions.
\end{abstract}


\maketitle

\section{Introduction}

There is a longstanding policy interest in unraveling how knowledge generated from public research is used in the private-sector. Studies towards this goal have heavily focused on patent data and considered citations between patents as evidence of knowledge flow. Despite some criticism \citep{Meyer-does-2000}, such notion has been widely accepted in the literature. Consequently, substantial attention has been paid to patents assigned to universities and other public organizations, examining how those patents are cited by other patents, especially by patents from companies \citep{Trajtenberg-univ-1997, Rosell-have-2009}.

University patents, however, only account for a small portion of granted patents, and the main products of public research are scholarly papers rather than patents. Just as patents, papers can also be cited by patents, and indeed both the cited patents and cited papers are served as the ``prior art'' of a patent application, playing a significant role for patent examiner to determine the patentability of the application. There has been a large literature on both the patent-to-patent and the patent-to-paper citation linkage. Yet, systematic studies, as we shall present in this paper, have been relatively scarce.

Our primary interest in this work is in the life science sector. The last several decades have seen an unprecedented rapid progress of life science, both in basic scientific discoveries and clinical medicine. Recent studies have suggested that biotechnology and pharmaceutical patents have been the main driver for the overall growth of patents and exhibit a particularly prominent ``science linkage'' \citep{Mowery-ivory-2004}. This has prompted us to ask: How has the patent-to-paper citation linkage of life science patents changed over time? In particular, we aim to answer the following lines of research questions:
\begin{enumerate}
\item How has the amount of science linkage changed over time? Does the change vary across different technology classes?
\item On the cited side of the linkage, which countries and types of institutions produce the cited papers? Whether basic or applied research are more likely to be cited?
\item On the citing side, to what extent company patents cite public science?
\end{enumerate}
These questions are important due to their high relevance to the policy community. Although the study of science linkage of patents has a long history, initiated by Narin and his colleagues in the 1980s \citep{Narin-status-1992, Narin-linkage-1997, McMillan-biotech-2000}, an up-to-date ``status report'' of science linkage has been lacking in the literature, partially due to the daunting task of resolving non-patent references to corresponding scholarly papers. Even in Narin's landmark study \citep{Narin-linkage-1997}, the analyzed patents were granted in two two-year periods (1987--1988 and 1993--1994). By contrast, our analysis covers patents from 1976 to 2012. Such a large-scale corpus allows us to probe how the science linkage has changed over time. By using a large sample over a 36-year period, we contribute to the literature a systematic accounting of linkage from technology to science. On the methodology side, we use a novel, paper-level indicator to quantify to what extent a paper is basic science or clinical medicine, allowing us to distill new insights on the science-technology linkage in biomedicine.

The rest of the paper is organized as follows. Section~\ref{sec:lit} discusses the context of our work. In Section~\ref{sec:data}, we describe the data source, selection of the cohort of patents analyzed in this work, and methods used to identify various properties of patents and cited papers. Section~\ref{sec:res} presents the results of our analysis. Finally, we discuss and conclude in Section~\ref{sec:dis}.

\section{Literature review}
\label{sec:lit}

This section briefly reviews three lines of literature that are closely related to our work. The first two are about knowledge flows as evidenced from patent-to-patent and patent-to-paper citations, and the third one presents some alternative interpretations other than knowledge flows.

\subsection{Knowledge flow as evidenced from patent-to-patent citations}

Many studies have compared the importance of patents, as operationalized as the number of citations it receives, from different sectors.
\citet{Jaffe-flows-1996} confirmed the geographic localization of
citations and found that university patents are cited more frequently and
government patents are cited less than company patents. \citet{Henderson-univ-1998} pointed out that the importance of university
patents has been overshadowed by
the increasing rate of university patenting. This finding was challenged later by \citet{Sampat-changes-2003} that found that such a decline is due
to the ``changes in the intertemporal distribution of citations to university
patents.'' \citet{Bacchiocchi-knowledge-2009} compared patent citations across countries and technological fields, showing that chemical, drugs \& medical, and
mechanical patents from U.S. universities are more cited than company patents, which does not hold for Europe and Japan patents. Numerous works have used regression frameworks to measure the likelihood of knowledge flow between patents assigned to different types of institutions, suggesting that university patents are more important than corporate ones in terms
of knowledge diffusion \citep{Jaffe-flows-1996, Trajtenberg-univ-1997,
Bacchiocchi-knowledge-2009}. \citet{Rosell-have-2009} found that for a limited
scope of technological fields, there was a more-than-half decline of both
knowledge inflows and outflows during the 1980s.

\subsection{Knowledge flow as evidenced from patent-to-paper citations}

We now shift our attention to the studies of patent-to-paper knowledge flow.
This topic has a longer history than that of the patent-to-patent case, dating back to the 1980s when Narin and his colleagues published a ``status report''
examining the time and nation dimensions of the science-technology
linkage \citep{Narin-status-1992}. Follow-up studies increased the timespan of
analyzed data and pointed out the increasingly heavy reliance of private-sector
patents on public science \citep{Narin-linkage-1997}. Particularly related to
our work is \citet{McMillan-biotech-2000} that concluded that the dependence of
biotechnology patents on public science is much heavier than other industries.

Several studies have found that patent-to-paper citations better represent knowledge flow when
comparing to patent-to-patent citations. For example, \citet{Lemley-examiner-2012} showed
that patent examiners generate less amount of NPRs. Later studies
reinforced this claim \citep{Roach-lens-2013}. \citet{Sorenson-science-2004}
found that patents that cite published non-patent literature have more
citations, implicating the important role of publications in technological
innovation.

Some works have examined the country dimension of patent-to-paper citations. \citet{Tijssen-global-2001} analyzed Dutch-authored papers
referenced in patents granted at the USPTO and found the dominance of
self-citation for domestic citation links. \citet{Acosta-science-2003}
uncovered significant differences between scientific citations in sectors and
patent citations in Spanish regions. \citet{Guan-china-2007} explored the
science-technology linkage in terms of regions and sectors for Chinese patents
at the USPTO and showed the heterogeneity in cited journals.

Other studies have instead paid attention to different technology sectors.
\citet{Popp-from-2017} analyzed three types of knowledge flow, namely
patent-to-patent, paper-to-paper, and patent-to-paper, in the alternative
energy sector and revealed that papers produced by government research are more
likely to accrue patent citations than any other types of institutions,
highlighting an important role of government research in translating from basic
to applied research. The analysis also emphasized a less important role for
universities in wind research, when compared to solar and biofuels research. \citet{Du-measure-2019} looked at the grant-publication-patent-drug linkage and observed that, among others, the vast majority of papers that are cited by drug patents are publicly funded.

NPRs have also been used to assist in the identification of novel patents.
\citet{Verhoeven-measuring-2016}, for example, measured novelty of patents in
terms of both combinatorial novelty of cited patents and novelty in knowledge
origin, which is based on NPRs.

\subsection{Interpretation of patent citations}

While the vast majority of literature interpreted patent citations as knowledge flow, some studies have criticized this
interpretation and proposed alternatives. \citet{Meyer-does-2000} looked at nanoscale patents and suggested that citation linkages from citing patents to cited papers hardly represent direct knowledge-transfer.
\citet{Callaert-source-2014} argued that patent-to-paper citations reveal the
relatedness between science and technology.
\citet{Fleming-science-2004} viewed inventions as combinatorial
search and hypothesized that science helps direct inventors' search process to
more useful combinations, therefore helping increase invention rate.

\subsection{Indicators for ``basicness'' of papers}

A repeatedly occurring assumption in the literature about the role of public
science is that public science institutions conduct basic science, while
private firms perform applied research by utilizing findings from basic
research. Yet, few studies have
examined to some extent papers cited in patents are basic research, possibly because of the difficulty in the
operationalization of the two notions to papers. One notable exception is
\citep{McMillan-biotech-2000} that used the four-level classification
scheme developed by the CHI Research in the 1970s \citep{Narin-structure-1976}. The scheme assigns journals to one of the four categories,
which are, from the most basic to the most applied, ``basic research'', ``clinical investigation'',
``clinical mix'', and ``clinical observation''. Focusing on the biomedicine domain, a recent proposal from \citet{Weber-identify-2013} used MeSH terms to develop an indicator of whether a paper is basic research or clinical research. The indicator is constructed based on whether the MeSH terms of a paper contain cell-, animal-, and human-related terms and classifies it as clinical research if there are human-related terms, in accordance with the widely adopted definition that clinical research is the study with human subjects.

\section{Data and Methods}
\label{sec:data}

\subsection{Sample selection}

The NBER patent database \citep{Hall-nber-2001} has been one of the major
sources for information about U.S. patents. However, it only covers patents
granted until 2006, whereas we want to extend to later patents. We therefore
used patent data directly from the USPTO and parsed the downloaded XML
files (\url{https://bulkdata.uspto.gov/}) to obtain
bibliographic information of patents. The NBER dataset instead is used as an
auxiliary source when we infer various attributes of patents.

As we are interested in science-technology linkage in the life science domain,
we need to select life science patents. In doing this, we note that there is an
inherent trade-off regarding sample coverage. On one hand, it may not be
desirable to narrow our analysis to, for example, patents about drugs that treat
certain diseases. On the other, selecting patents from other domains, such as
the software industry, may bias our statistics about science linkage, since those
patents seldom cite biomedical papers. Here we leverage the categorization developed by NBER \citep{Hall-nber-2001}, which segments patents into six categories. We define life science patents as those belonging to one of the two NBER technological categories, namely Chemical
(Category 1) and Drugs \& Medical (Category 3). Operationally, we selected
not-withdrawn (\url{https://www.uspto.gov/patents-application-process/patent-search/withdrawn-patent-numbers}),
utility patents granted between 1976 and 2012 whose primary, three-digit USPC
(U.S. Patent Classification) technology codes are in the 92 codes corresponding
to the two NBER categories (Appendix 1 in \citep{Hall-nber-2001}). The final
sample used in our study consists of $1,088,650$ patents. Patents that are not included into our sample are from the following NBER categories: Computers \&
Communications (Category 2), Electrical \& Electronic (Category 4), Mechanical (Category 5), and Others (Category 6).

\begin{table}
\centering
\caption{Top 20 countries with most patents.}
\label{tab:top-country}
\begin{tabular}{c r r || c r r}
Country & Patents & \% & Country & Patents & \% \\ \hline
US & 602695.19 & 55.42 & TW & 9905.90 & 0.91 \\
JP & 156946.13 & 14.43 & SE & 9244.52 & 0.85 \\
DE & 87173.10  & 8.02 & BE & 7976.88 & 0.73 \\
FR & 35592.16  & 3.27 & IL & 6849.96 & 0.63 \\
GB & 33518.90  & 3.08 & AU & 6410.51 & 0.59 \\
CA & 20636.18  & 1.90 & DT & 5365.60 & 0.49 \\
CH & 18003.52  & 1.66 & DK & 4885.43 & 0.45 \\
KR & 14994.77  & 1.38 & JA & 4850.13 & 0.45 \\
IT & 14563.86  & 1.34 & FI & 4075.29 & 0.37 \\
NL & 11000.00  & 1.01 & AT & 3522.86 & 0.32 \\
\end{tabular}
\end{table}

\subsection{Country origin of patent}

To examine whether science linkage varies across patents from different countries, we need to identify the country origin of a patent. We do so by looking at the residences of all inventors. For a tiny portion (0.63\%) of patents whose country of origin cannot be determined through this way, which is due to missing data of the first inventor's address, we use the NBER data to locate the country.

Table~\ref{tab:top-country} lists the top 20 countries that have the largest
number of patents, based on fractional counting. They in total contribute to 97.3\% of all patents in our
cohort. It is clear that the US patent system has granted life science patents
from inventors originated from diverse countries, although US accounts for
more than half of the patents. These country statistics remain very similar if we use the residence of the first-inventor to identify country origin (Appendix A).

\subsection{Type of patent assignee}

To study how the types of assignees may affect citations to scientific papers, we need to classify patent assignees. To do this, we again leverage the NBER patent dataset that has already classified assignees of patents in 1976--2006 into one of the
following six types: corporation, university, institute, government, hospital,
and individual. For later patents, we assign the type based on the exact match
of assignee names. If this fails, we then classify by checking the role of the
assignee provided by USPTO, whether the assignee name is the same as an applicant,
and whether it contains certain keywords (e.g., ``Ltd'', ``University''). There
are 9.76\% patents without any assignee listed.

\begin{figure}[t]
\centering
\includegraphics[trim=0mm 8mm 0mm 0mm, width=\columnwidth]{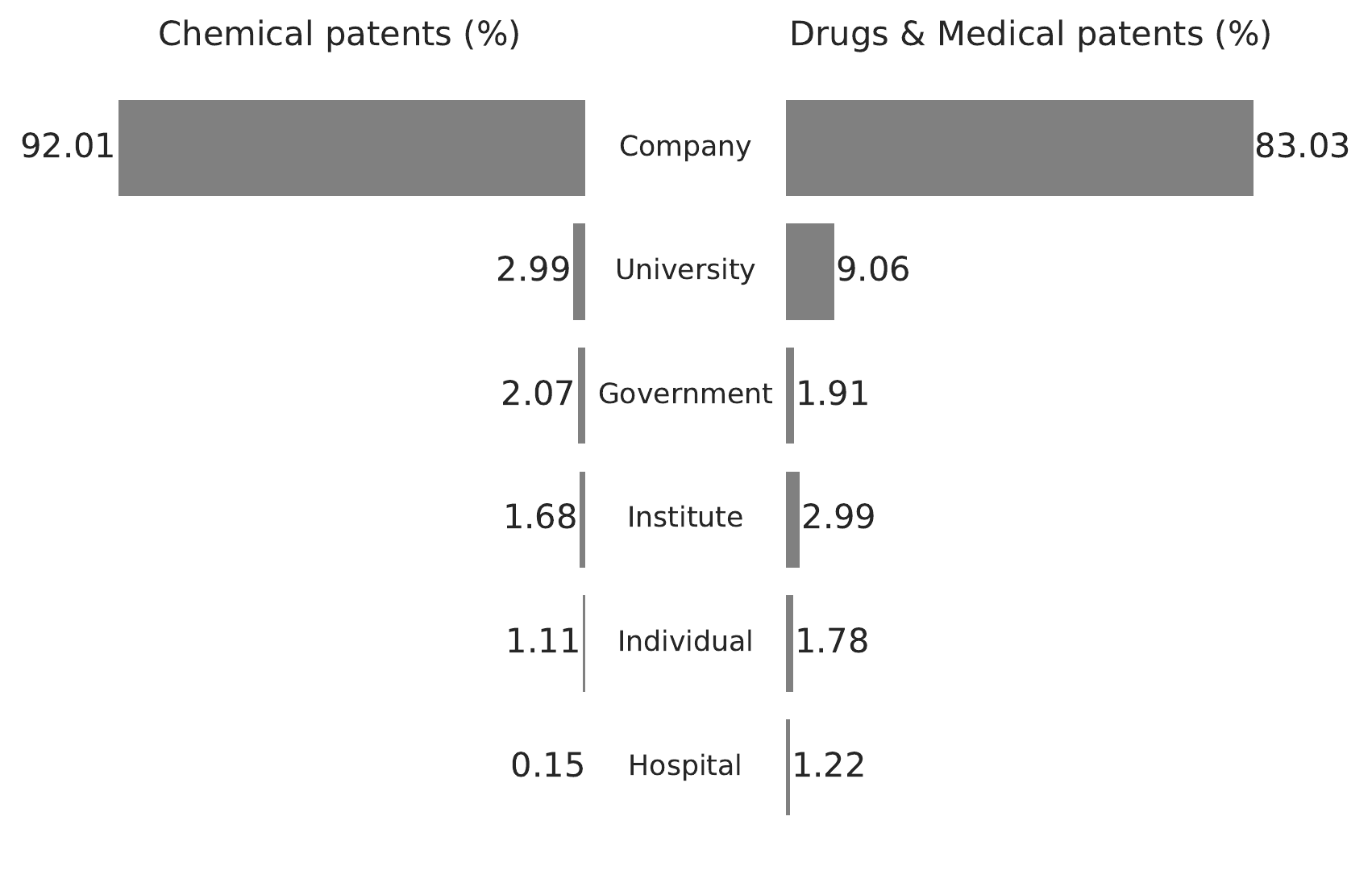}
\caption{Distribution of types of assignees.}
\label{fig:assg-dist}
\end{figure}

Figure~\ref{fig:assg-dist} gives a side-by-side comparison of the decomposition
of the types of assignees for both chemical and drugs \& medical (DM) patents. Not surprisingly,
the overwhelmingly majority of patents are assigned to companies. A 
larger fraction of DM patents, however, come from universities. Many previous works
have linked this to the Bayh--Dole Act that permits universities to own
inventions that are funded by government \citep{Mowery-Bayh-2004}.

\subsection{Non-patent references in patent}

Each and every NPR cited in the patents has been resolved previously to determine whether and which MEDLINE paper it refers
to, with a high accuracy obtained \citep{Ke-compare-2018}. MEDLINE is perhaps the most widely used database for the biomedical research literature, curated and maintained by the US National
Library of Medicine (NLM). It is publicly available and provides a variety of meta data about papers indexed there, including common bibliographic information like authors, affiliations, journal, publication year, funding, etc. It also provides domain specific information like Medical Subject Headings (MeSH). Moreover, many additional resources that we rely on
have been built on top of MEDLINE, and literature has been using MEDLINE for innovation study, such as operationalization of the triple-helix model based on MeSH terms \citep{Petersen-triple-2016}.

\subsection{Country and institution type of papers}

To understand how public science contributes to knowledge cited in patents, we 
need to classify the types of institutions of papers. However, an important question 
before the classification is which author's (or authors') affiliation we should
use, as modern science has become a collaboration
endeavor \citep{Wuchty-team-2007}. Here we choose to look at only the
first-author's affiliation for two reasons. First, as stated from the NLM, ``until 2014, only the affiliation of the first author was
included,'' (\url{https://www.nlm.nih.gov/bsd/mms/medlineelements.html\#ad}) and the first author's affiliation was not included until 1988.
This limitation is also reflected in the data: 87\% of the $218,483$ papers cited in
patents and without author affiliations were published before 1988. Second, in
biomedical research, it is generally accepted that the first and the last author
get the most credit of a paper for performing and supervising the research,
respectively, and the two authors share the same affiliations in most cases.

From the text of the first author's affiliation, we seek to extract the country and
institution type information. This task, fortunately, has been fulfilled by an
online tool called
\textsc{MapAffil} (\url{http://abel.lis.illinois.edu/cgi-bin/mapaffil/search.py}; \citep{Torvik-MapAffil-2015}). It returns geography information and institution type of the input MEDLINE paper and has a reported accuracy of 97.7\%. \textsc{MapAffil} classifies institutions into eight categories, namely educational, hospital,
educational hospital, organization, commercial, government, military, and unknown.
For our study, we merge educational hospital into educational, since teaching
hospitals still serve the education role for training medical students. Furthermore,
we combine the organization, government, and military categories into a single one,
called public research organization (PRO), because we primarily concern about
whether or not cited research are performed by companies. Previous studies have
also employed a similar procedure \citep{Bacchiocchi-knowledge-2009}. Therefore,
there are five types of institutions of papers, namely educational (EDU), PRO, hospital (HOS), commercial (COM), and unknown (UNK).

\subsection{Funding support of papers}

An ongoing effort in the study of the patent-to-paper citation linkage is to understand to what extent cited papers are supported by public funding. We retrieve this information from the paper meta data provided in the MEDLINE database. First, we determine whether a paper is funded by the US government by looking at whether the ``Publication Type'' field has any of the following four terms: ``Research Support, U.S. Gov't, Non-P.H.S.'', ``Research Support, U.S. Gov't, P.H.S.'', ``Research Support, N.I.H., Extramural'', and ``Research Support, N.I.H., Intramural''. Second, we determine whether a paper is supported by the NIH by looking at the ``Grant List'' field and further record which NIH institutes support the paper.

\subsection{``Basicness'' of papers}

In this study, we do not adopt the method proposed in \citet{Narin-structure-1976} to quantify ``basicness'' of papers for four reasons. First,
we are not aware the scheme is publicly available. Second, it remains unclear whether a scheme developed in the 1970s is still applicable nowadays, with numerous journals established since then. Third, the scheme only considers journals
indexed in the SCI, while many MEDLINE journals are not there. Lastly and most importantly, the scheme operates
on journals rather than papers. One immediate implication of this is that papers
published in all the journals belonging to the same category have the same ``basicness.'' This is problematic,
because many biomedical journals publish qualitatively different types of research, which can be basic or applied. As an example, \emph{Circulation}, a prestige journal with a 2017 Impact Factor of 18.88, ``publishes [...] related to cardiovascular health and disease, including observational studies, clinical trials, epidemiology, health services and outcomes studies, and advances in basic and translational research'' (\url{https://www.ahajournals.org/circ/about}).

Here we use an innovative method that was recently proposed to identify translational
science in biomedicine \citep{Ke-identify-2019}. Translation science is research that ``translate'' basic scientific discoveries (bench-side or basic research) to clinical
applications (bed-side or applied research). The method quantifies the basicness of papers directly. It results in
a paper-level indicator called level score (LS) ranging from -1 to 1, with LS closer to -1 meaning that the paper is, by construction, more basic and 1 more applied. The method learns similarities between MeSH terms based on their co-occurrences among papers, using modern representation learning techniques. It then identifies an axis that points from basic science terms to applied ones. MeSH terms are organized into a hierarchical structure, and each of them has a location in the tree. For example, ``Eukaryota'' (B01) is within branch B (``Organisms''). Given this tree, a MeSH term is a basic science one if it is located within the following terms: ``Cells'' (A11), ``Archaea'' (B02), ``Bacteria'' (B03), ``Viruses'' (B04), ``Molecular Structure'' (G02.111.570), ``Chemical Processes'' (G02.149), and ``Eukaryota'' (B01) except ``Humans''. A term is applied if it is located within the following nodes: ``Humans'' (B01.050.150.900.649.801.400.112.400.400) and ``Persons'' (M01). The basicness of a MeSH term is its projected position onto the axis, expressed as the cosine similarity between the axis vector and the term vector. The LS of a paper is the average basicness of its MeSH terms. The method has been validated and is consistent with Narin's four-level classification and other existing methods.

\begin{table*}[t!]
\centering
\caption{Summary statistics of non-patent references (NPRs) cited by U.S. life
science utility patents 1976--2012, grouped by their NBER categories. MNPR
refers to an NPR that corresponds to a MEDLINE paper.}
\label{tab:summary}
\begin{tabular}{c c | r r r | r r r | r r}
\hline \hline
\multicolumn{10}{c}{NBER Category 1: \emph{Chemical}} \\
& & \multicolumn{3}{c|}{Patents} & \multicolumn{3}{c|}{Total} & \multicolumn{2}{c}{Mean MNPRs by} \\
Sub-cat & Name                        & All       & w/ MNPRs & \%    & NPRs        & MNPRs     & \%   & All   & w/ MNPRs \\ \hline
11      & Agriculture, food, textiles & $22,166$  & $1,019$  & 4.60  & $54,183$    & $2,853$   & 5.26  & 0.129 & 2.800 \\
12      & Coating                     & $58,326$  & $1,873$  & 3.21  & $127,440$   & $9,402$   & 7.38  & 0.161 & 5.020 \\
13      & Gas                         & $20,196$  & $319$    & 1.58  & $32,179$    & $1,350$   & 4.20  & 0.067 & 4.232 \\
14      & Organic compounds           & $91,301$  & $26,538$ & 29.07 & $686,384$   & $291,540$ & 42.47 & 3.193 & 10.986 \\
15      & Resins                      & $105,960$ & $15,667$ & 14.78 & $585,437$   & $288,730$ & 49.32 & 2.725 & 18.429 \\
19      & Miscellaneous               & $384,434$ & $20,049$ & 5.22  & $827,008$   & $123,405$ & 14.92 & 0.321 & 6.155 \\
& \multicolumn{1}{r|}{\emph{Total:}}  & $682,383$ & $65,465$ & 9.59  & $2,312,631$ & $717,280$ & 31.02 & 1.051 & 10.957 \\
\\
\multicolumn{10}{c}{NBER Category 3: \emph{Drugs $\&$ Medical}} \\
& & \multicolumn{3}{c|}{Patents} & \multicolumn{3}{c|}{Total} & \multicolumn{2}{c}{Mean MNPRs by} \\
Sub-cat & Name                           & All       & w/ MNPRs  & \%    & NPRs        & MNPRs       & \%    & All    & w/ MNPRs \\ \hline
31      & Drugs                          & $158,665$ & $89,008$  & 56.10 & $2,225,049$ & $1,395,016$ & 62.70 & 8.792  & 15.673 \\
32      & Surgery \& medical instruments & $137,981$ & $28,975$  & 21.00 & $668,424$   & $274,526$   & 41.07 & 1.990  & 9.474 \\
33      & Biotechnology                  & $79,148$  & $63,625$  & 80.39 & $1,618,241$ & $1,141,578$ & 70.54 & 14.423 & 17.942 \\
39      & Miscellaneous                  & $30,473$  & $5,748$   & 18.86 & $123,833$   & $46,377$    & 37.45 & 1.522  & 8.068 \\
& \multicolumn{1}{r|}{\emph{Total:}}     & $406,267$ & $187,356$ & 46.12 & $4,635,547$ & $2,857,497$ & 61.64 & 7.034  & 15.252 \\ 
\hline \hline
\end{tabular}
\end{table*}

\section{Results} \label{sec:res}

\subsection{Summary statistics}

Table~\ref{tab:summary} reports the overall statistics of NPRs cited in the
$1,088,650$ patents in our sample, grouped by their NBER subcategories. The first group of statistics in Table~\ref{tab:summary} concerns about the total number of patents. Chemical patents share 62.7\%, and the rest are DM patents. Among chemical patents, resins and organic compounds are the two largest subcategories, whereas drug and surgery \& medical instruments patents are most presented ones in the DM category. Overall only $252,821$ (23.2\%) patents have at least one NPR linked to a MEDLINE paper (hereafter MNPR). This fraction, however, varies significantly across the two categories: only 9.6\% for chemical patents but 46.1\% for DM ones. The variability also holds at the subcategory level. 29.1\% of resins patents and 14.8\% organic compounds patents have MNPRs; by contrast, 80\% of biotechnology patents cite MEDLINE papers, while 56.1\% of drugs patents and 21\% of surgery \& medical instruments patents do so.

The second group of statistics is the total number of NPRs and MNPRs. A total of $6,948,178$ NPRs were emanated from our corpus of patents, among which $2,312,621$ (33.3\%) are from chemical ones. More than half ($3,574,777$; 51.4\%) of the NPRs are MNPRs, which are dominated by DM patents ($2,857,497$; 79.9\%). The rest ($717,280$; 20.1\%) are from chemical patents. Therefore, 
although there is a larger portion of chemical patents, they generate less amount of NPRs and are less linked to
science, when comparing to DM patents. As for the subcategories, 49.3\% and 42.5\% of NPRs in resins and organic compounds patents, respectively, refer to MEDLINE papers. Biotechnology and drug patents account for 88.8\% of all the MNPRs in the DM category, and 70.5\% and 62.7\% of their NPRs are scientific.

The last group of statistics is the average number of MNPRs per patent. Here we average by both all patents and patents with at least one MNPR, since the majority of patents have no MNPRs. On average, a chemical patent cites one MEDLINE paper; a DM patent cites 7 papers. Such contrast, however, is much less evident if we use the second averaging procedure. For patents that have at least one MNPR, a chemical patent has 11 MNPRs while 15 for a DM patent. Delving into subcategories, there is a large variation of the extent of linkage to science. Organic compounds and resin patents on average cite 3.2 and 2.7 MNPRs respectively. A biotechnology patent has on average 14 MNPRs, larger than any other categories.

In summary, all the overall statistics suggest that there is a huge variation of the volume of science linkage, which is dominated by biotechnology and drug patents. This result is consistent with a previous small-scale study \citep{McMillan-biotech-2000}.

\subsection{Overall characteristics over time}

\begin{figure*}[t]
\centering
\includegraphics[trim=0mm 2mm 0mm 0mm, width=\textwidth]{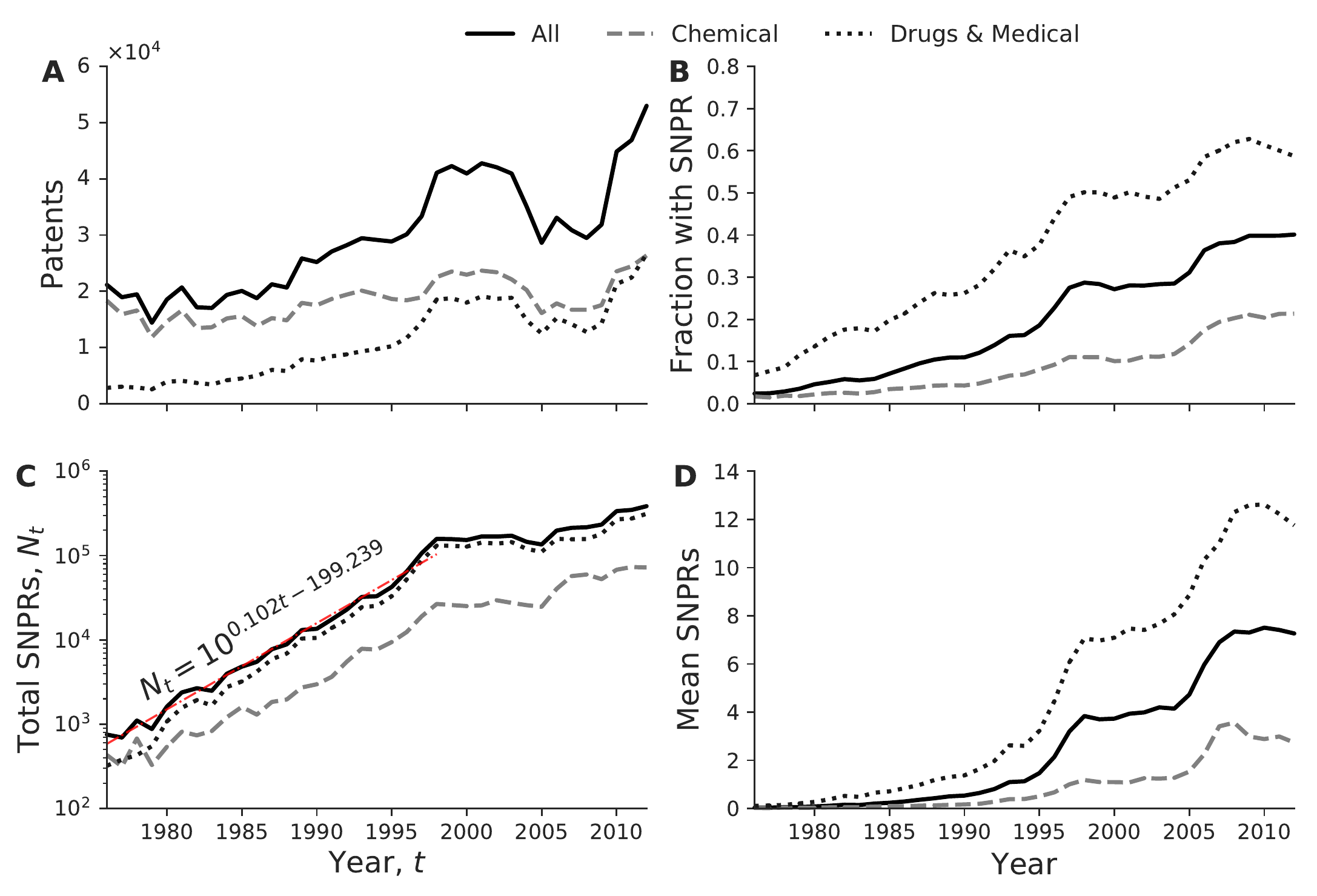}
\caption{Overall characteristics of the patent-to-paper citation linkage over
time. (A) The number of patents. (B) The fraction of patents with at least
one MNPR. (C) The
total number of MNPRs. The red, dash-dotted line represent an exponential fit of the total number of MNPRs from 1976 to 1998, $N_t = 10^{0.102 \cdot t - 199.239}$. (D) The average number of MNPRs per patent.}
\label{fig:overall}
\end{figure*}

Next, we investigate how overall characteristics change over time.
Figure~\ref{fig:overall}A shows a steady increase of the total number of granted
patents over the examined period, reaching from $21,151$ in 1976 to $52,994$ in 2012.
Such increase is largely driven by the remarkable growth of DM patents: a
nearly ten-fold increase from only $2,827$ in 1976 to $26,616$ in 2012. The
number of chemical patents, on the other hand, has increased relatively
slowly---44\%. We notice that there is a flatten period followed by a decreasing period from 1998 to 2005, for both chemical and DM patents. Accompanying the increase of the raw number of patents is an increasing fraction of
patents that cite MEDLINE-indexed papers, as presented in
Figure~\ref{fig:overall}B. In 1976, only 1.7\% chemical and 6.8\% DM patents had MNPRs,
and in 2012 the number reached to 21.3\% and 58.7\%, respectively. Figure~\ref{fig:overall}C plots the total number of 
patent-to-paper citations for patents granted in each year, demonstrating a remarkable increase of science linkage. We fit the growth from 1976 to 1998, obtaining $N_t = 10^{0.102 \cdot t - 199.239}$, where $t$ is the calendar year and $N_t$ is total citations at $t$. This means that there is an exponential growth of
the total number of MNPRs, which doubled every $\log_{10} 2 / 0.102 = 2.94$ years. DM patents, again, drive the increase, and generate more MNPRs than chemical patents across years. Finally, the increase of the total number of MNPRs is not due to the increase of the number of patents, but rooted at patents themselves, as confirmed in Figure~\ref{fig:overall}D which shows that the average number of MNPRs per patent also increases. Yet, DM patents have a faster increase than chemical patents.

\begin{figure*}[t]
\centering
\includegraphics[trim=0mm 2mm 0mm 0mm, width=\textwidth]{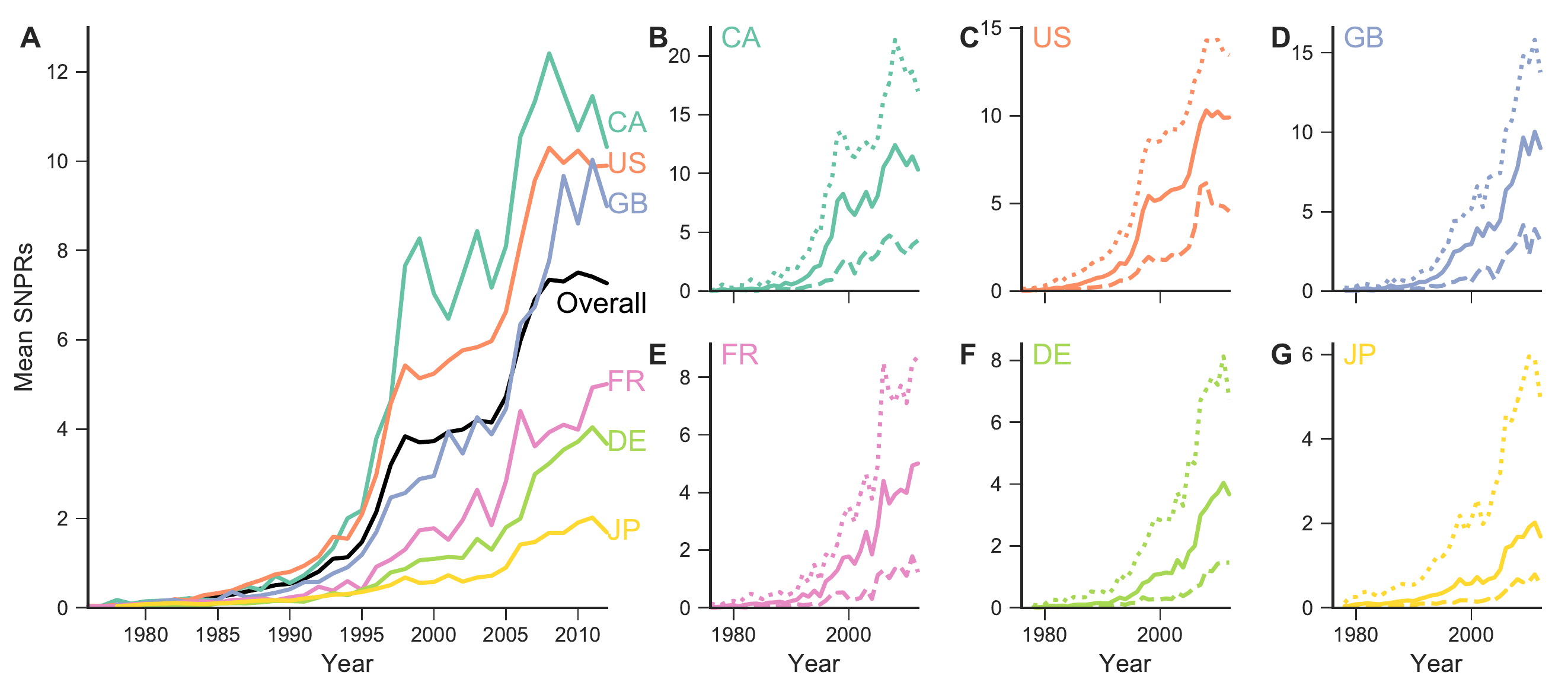}
\caption{The country dimension of the patent-to-paper citation linkage. (A) The
average number of MNPRs per patent over all patents and over patents from the six most-patented countries. (B--G) The average number of
MNPRs in patents originated from (B) Canada, (C) the U.S., (D) the United Kingdom,
(E) France, (F) Germany, and (G) Japan. Solid lines in (B--D) represent all patents in
the country, dashed lines chemical patents, and dotted lines drugs and medical
patents.}
\label{fig:country}
\end{figure*}

We then add the country dimension to the analysis of patent-to-paper citations.
Figure~\ref{fig:country}A shows that the average number of MNPRs per patent has
been increasing for patents originated from the top 6 countries with the largest number of
patents. The extent, however, varies by countries. For patents from Canada, the U.S., and the U.K., the average increases faster than the overall case, while for patents from France, Germany, and Japan, it increases slower than the overall case.
What is noteworthy is that, starting from around 1996, Canada has surpassed US in generating more MNPRs per patent.
Figures~\ref{fig:country}B--G further look at chemical and DM patents separately for each of the top 6 countries. From these figures, we can
conclude that (1) across the top countries and categories, there is an increasing citation linkage
from life science patents to biomedical research; and (2) DM patents exhibit a faster increase than chemical patents across years and countries.

\subsection{Cited science}

In this section, we explore the characteristics of papers that are cited by patents. We do so at the reference level; that is, a paper that is cited by multiple patents is counted multiple times, since the number of citations a paper receives from patents displays a heavy-tailed distribution, similar to the case of citations from papers \citep{Ke-compare-2018}.

\begin{figure*}[t]
\centering
\includegraphics[trim=0mm 2mm 0mm 0mm, width=\textwidth]{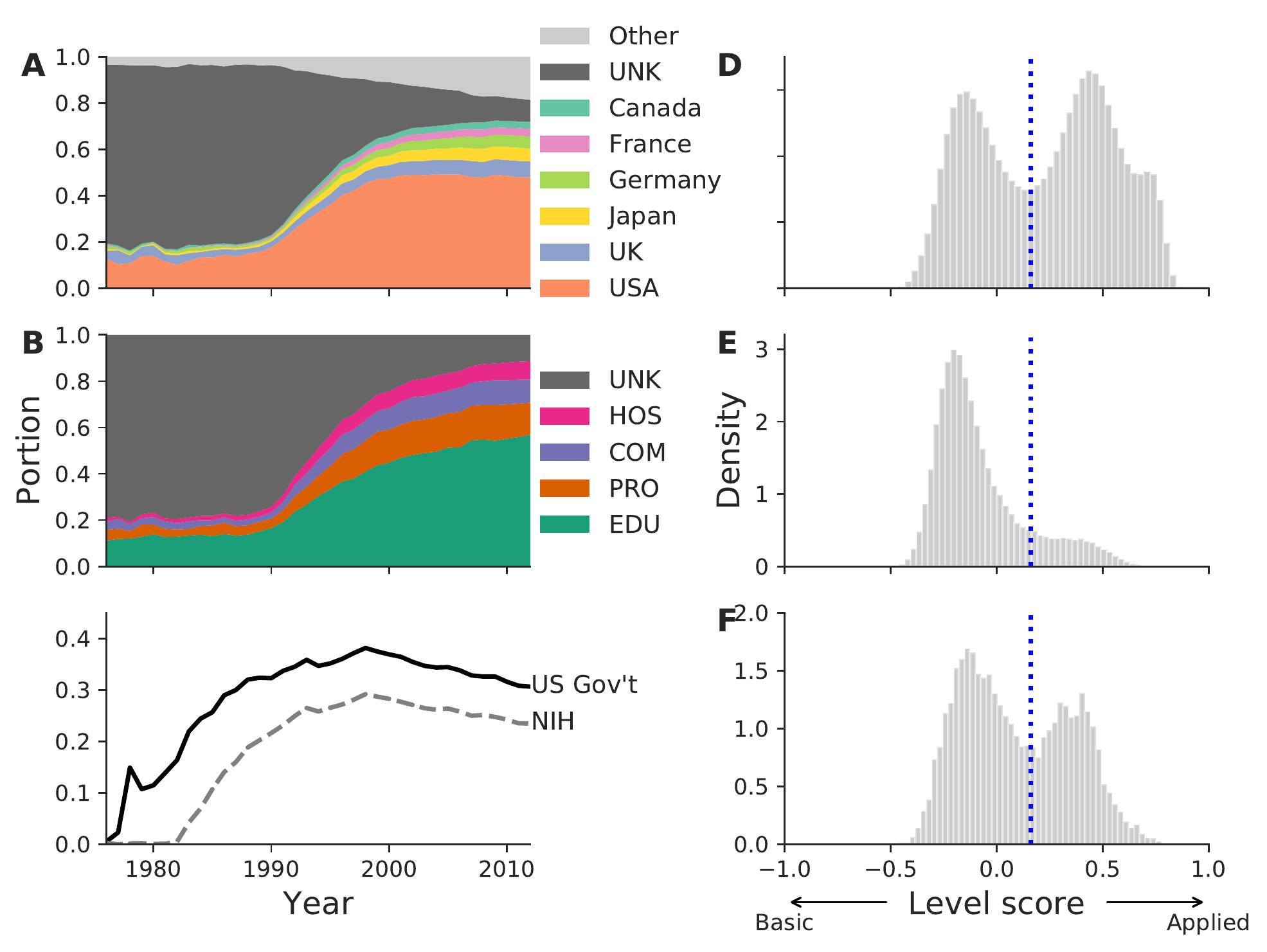}
\caption{Characteristics of cited science. (A--C) Fraction of MNPRs produced (A) by countries, (B) by institution types, and (C) supported by the U.S. government and the NIH in particular. (D) Histogram of level score for all
MEDLINE-indexed papers published between 1980 and 2012. The blue dotted line
indicates the score (0.16) corresponding to the local minimum of density. (E)
Histogram of level score of references cited in USPTO-issued patents. (F) The same as (E) but based on patents associated with FDA-approved
drugs. For 42.7\% of papers (D) and 85.2\% (E) and 59.6\% (F) of references,
the score is smaller than 0.16.}
\label{fig:cited-sci}
\end{figure*}

First, we study the distributions of countries where cited papers are produced. Figure~\ref{fig:cited-sci}A plots the fractions of MNPRs authored by different countries over time. Here we display the results separately for the six individual countries that have the largest shares at 2012 and combine the shares of the rest countries together. The distribution at a particular year is derived as follows. We first get all the patents granted in that year, and then count the number of MNPRs produced by a given country and normalize it by the total number of MNPRs cited by all the patents in that year. 

Figure~\ref{fig:cited-sci}A shows that the U.S. has been consistently the largest
producer of cited science, accounting for almost half (49\%) of the MNPRs cited by patents in 2012. Other top countries contribute to
significantly smaller fractions: 6.8\%
for the UK and 5.5\% for Japan. Note that here one may refrain to conclude that US
science has been increasingly cited by patents over time, because the apparent increase of the fraction of US science could simply due to an increasing portion of cited papers with affiliation information available. This is corroborated by the observation that the share of US science has been stable since around 2000.

Figure~\ref{fig:cited-sci}B presents the fractions of cited references that are
produced by different types of institutions over time, derived using the same procedure described above. Universities have been consistently
the largest producer; 57.7\% of references that are cited by patents granted in 2012 are written by them. PRO, which includes institutes and government, are the second major player, contributing to 9.8\%. Public science, therefore, share 67.5\% of cited science in patents. Companies account for only 10\%.

We also examine what are the funding agency that supported the science cited by patents. Figure~\ref{fig:cited-sci}C shows the portion of references that are supported by U.S. government and by NIH specifically. Since 1990, more than
30\% of cited science are supported by U.S. government and 20\% by NIH.
Table~\ref{tab:top-ic} further shows the top NIH institutes by the amount of citations they receive. 

\begin{table*}[t]
\centering
\caption{Number of citations for top NIH IC.}
\label{tab:top-ic}
\begin{tabular}{l r r}
IC & Citations & \% \\ \hline
National Cancer Institute & $416,642$ & 22.3 \\
National Institute of General Medical Sciences & $251,171$ & 13.5 \\
National Institute of Allergy and Infectious Diseases & $248,842$ & 13.3 \\
National Heart, Lung, and Blood Institute & $239,139$ & 12.8 \\
National Institute of Diabetes and Digestive and Kidney Diseases & $128,801$ & 6.9 \\
National Institute of Neurological Disorders and Stroke & $89,904$ & 4.8 \\
National Institute of Child Health and Human Development & $55,184$ & 3.0 \\
National Center for Research Resources & $54,456$ & 2.9 \\
National Institute on Aging & $49,538$ & 2.7 \\
National Institute of Diabetes and Digestive and Kidney Diseases & $40,886$ & 2.2 \\
\end{tabular}
\end{table*}

The last effort to characterize the cited science is to examine to what extent
they are basic or applied research. We use the LS indicator described in Section~\ref{sec:data} to measure the basicness of each paper. First, we plot in Figure~\ref{fig:cited-sci}D the histogram of LS for all the $14,916,511$ MEDLINE papers published between 1980 and 2012, illustrating how the entire biomedical literature is distributed
along the basic-applied spectrum. This serves as the baseline set, against which we compare with the set of papers cited by patents. We observe a bimodal distribution.
Figure~\ref{fig:cited-sci}E shows the same plot, but for the references that are
cited by patents.
We clearly observe that the vast majority of these MNPRs situate at the basic end. As a simple calibration, the bimodality in Figure~\ref{fig:cited-sci}D allows us to empirically find a threshold $th$ to separate the two modes, which is $0.16$. For 42.7\% of all papers in Figure~\ref{fig:cited-sci}D, their score is smaller than $th$; By contrast, $85.2\%$ of MNPRS in Figure~\ref{fig:cited-sci}E fall into this category. This result is robust if we instead look at the paper level. We further make additional measurements to ascertain that the observation is
not driven by patents with many MNPRs. For each patent, we calculate (1) the
average value of LS of its cited papers; and (2) the fraction of papers with LS
smaller than $th$. The results confirm that for the vast majority of patents,
most of their references are papers from the basic side.

As a separate case study, we examine MNPRs from patents that are associated with
drugs approved by the U.S. Food and Drug Administration (FDA). The Hatch--Waxman Act mandates that drug innovators to provide FDA with the list of patents that covers the drug, and FDA included these patents in the \emph{Approved Drug Products With Therapeutic Equivalence Evaluations} (also known as the Orange Book), although it is not FDA's task to actually evaluate the coverage. Such patents may possess economic value for their owner
to surpass the cost of the development of drugs, and at the same time have the
cure value for patients. We get this list of patents from \url{https://www.fda.gov/drugs/informationondrugs/ucm129689.htm}. We find a much smaller number ($4,380$) of such
patents, which cite $28,512$ MNPRs in our sample of papers.
Figure~\ref{fig:cited-sci}F shows that, although most (59.6\%) of
these MNPRs are on the basic side, substantial amount are on the applied side,
yielding a bimodal distribution that is not present in the overall case in Figure~\ref{fig:cited-sci}E. This
may be related to the underlying process of drug development where
pharmaceutical companies need to test the safety and effectiveness of drugs on human---which is applied research by definition.

\subsection{Private-sector patents}

In this section, we analyze science linkage of patents assigned to companies. Table~\ref{tab:com} presents the overall percentages of citations originated
from company patents to papers authored by different types of affiliations. We observe that about 48\% of citations form company patents go to university papers, and this varies little if we focus on chemical or DM patents separately or US patents only. Other public research organization
ranks the second, contributing to 13--15\%. Companies share only 11--13\% of the science base of their
patents.

\begin{table}[t]
\centering
\caption{Percentage of MNPRs originated from companies to different types of
institutions.}
\label{tab:com}
\begin{tabular}{c | c c | c c | c c}
& \multicolumn{2}{c|}{All} & \multicolumn{2}{c|}{Chemical} & \multicolumn{2}{c}{DM} \\
& All & US & All & US & All & US \\ \hline
EDU & 48.1 & 47.6 & 47.7 & 47.7 & 48.2 & 47.6 \\
PRO & 13.4 & 13.0 & 14.8 & 14.5 & 13.1 & 12.6 \\
COM & 11.5 & 11.6 & 13.1 & 13.2 & 11.1 & 11.2 \\
HOS & 7.1 & 7.4 & 5.4 & 5.6 & 7.5 & 7.9 \\
UNK & 19.9 & 20.4 & 19.0 & 19.0 & 20.1 & 20.8 \\
\end{tabular}
\end{table}

\begin{figure}[t]
\centering
\includegraphics[trim=0mm 2mm 0mm 0mm, width=\columnwidth]{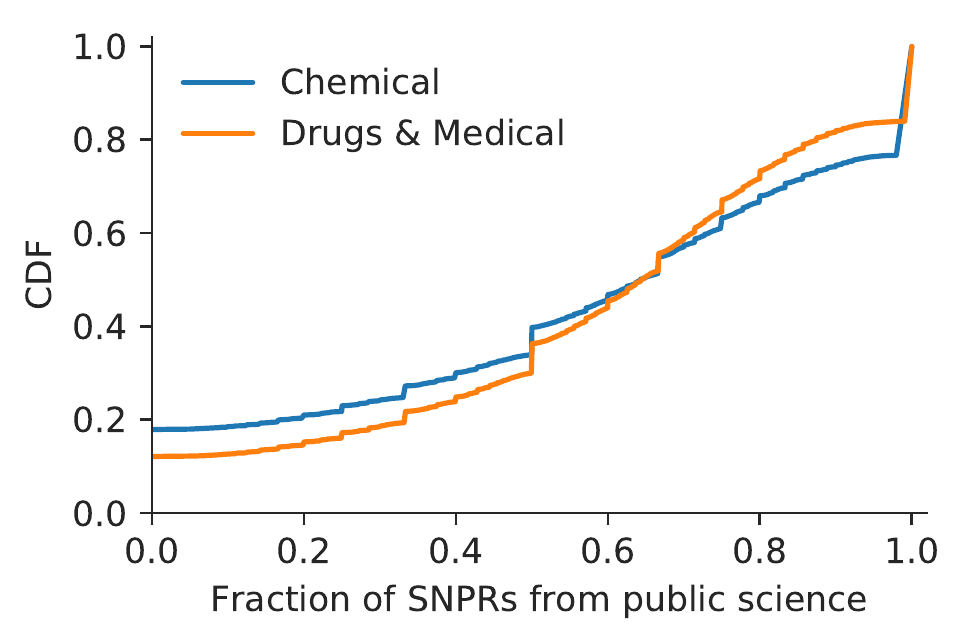}
\caption{Cumulative distributions of fraction of public science MNPRs cited in company patents.}
\label{fig:com}
\end{figure}

We further look at the science linkage of individual companies. By way of example, Medtronic, a global medical device company, owns the largest number ($3,565$) of DM patents in our sample. We find $11,242$ MNPRs in those patents, among which $5,824$ are from universities, 579 from PRO, and only 297 from companies. The fraction of MNPRs authored by the public science section (universities and PRO), therefore, is 0.57. Table~\ref{tab:top-com} extends this calculation to the top 10 companies that have the largest number of chemical and DM patents, indicating a significant linkage to public science. We make one more step and repeat this calculation to all the companies whose patents have at least one MNPR. Figure~\ref{fig:com} shows the cumulative distributions of fraction of public science MNPRs for all those companies, across the chemical and DM categories. For more than 60\% of companies, more than half of MNPRs cited in their patents are from public science.

\begin{table*}[t!]
\centering
\caption{The top 10 companies that own the largest number of chemical (top) and DM (bottom) patents. The Fraction columns refer to the fraction of MNPRs that are authored by the public science section (universities and PRO).}
\label{tab:top-com}
\begin{tabular}{l r r}
Company & Patents & Fraction \\ \hline
\multicolumn{3}{c}{Chemical} \\
BASF AG & $8,523$ & 0.57 \\
Bayer AG & $8,450$ & 0.42 \\
E. I. du Pont de Nemours and Company & $7,462$ & 0.58 \\
General Electric Company & $7,276$ & 0.73 \\
Eastman Kodak Company & $7,028$ & 0.46 \\
Fuji Photo Film Co., Ltd. & $6,463$ & 0.56 \\
The Dow Chemical Company & $5,545$ & 0.47 \\
Ciba-Geigy Corporation & $5,059$ & 0.31 \\
Hoechst AG & $4,468$ & 0.36 \\
Shell Oil Company & $4,076$ & 0.73 \\
\multicolumn{3}{c}{Drug \& Medical} \\
Medtronic Inc. & $3,565$ & 0.57 \\
Merck \& Co., Inc. & $3,000$ & 0.47 \\
The Procter \& Gamble Company  & $2,349$ & 0.52 \\
Eli Lilly and Company & $2,314$ & 0.42 \\
Bayer AG & $2,258$ & 0.54 \\
Pioneer Hi-Bred International, Inc. & $2,064$ & 0.81 \\
Cardiac Pacemakers, Inc. & $1,955$ & 0.61 \\
Pfizer Inc. & $1,816$ & 0.55 \\
Abbott Laboratories & $1,736$ & 0.58 \\
Monsanto Technology LLC & $1,696$ & 0.80 \\
\end{tabular}
\end{table*}

\section{Discussion}
\label{sec:dis}

We have uncovered several empirical findings regarding how science linkage of US life science patents has changed over a 37-year period. From the prevalent perspective of viewing citation linkage as knowledge flow, this study is particularly important, because our results suggest a continuous linkage of public science to private sector inventions. First, the overall growth of life science patents are largely driven by the increase of drug and medical patents. The volume of science linkage are increasing exponentially, doubling every 2.9 years. The increase happens in both chemical and drugs \& medical patents, as well as patents originated from different countries. 

Second, almost half of the MNPRs are produced in the US; the majority of them are from the public science sector. Public science---research performed by academics and government institutions---is widely acknowledged to have a strong influence on technology development. Our work provides empirical, quantitative, and longitudinal evidence of the magnitude of the dependence of technologies on public science.

Third, the overwhelming of cited science are basic research; yet, the nuance is for patents associated with drugs, with a non-negligible portion of them are applied research. The premise that basic science lays foundations for applied science is extensively discussed and widely embedded in many theoretical models about science-technology interaction (\emph{e.g.}, the ``linear model'' of innovation running from basic science to applied science to technologies and economic growth \citep{Balconi-defence-2010}). Our work pushes forward this line of inquiry, by moving from a dichotomous question of whether basic science fuels applied research towards a quantitative understanding of the extent of the reliance. Such a complication is important, because our findings in general corroborate the pivotal role of basic science, but at the same time point to a previously ignored contribution of applied research. On this regard, our work supports \citet{Gittelman-revolution-2016}, which argues that understanding diseases requires embedding applied science into basic research.

Fourth, the US government and NIH in particular continue to be found as funders of research cited in patents. Last, for the majority of companies, most of their patents cite public science. However, to what extent the linkage represents a direct knowledge flow is a line of challenging future work.

Many previous works about the role of public science in private sector innovation assumed that public research organizations conduct basic research, while private firms perform applied research by utilizing results generated from basic research. However, few studies have examined whether papers that are cited in patents are basic research. We bridge this gap by using a novel indicator proposed in our previous work that quantifies the extent to which a paper is basic research or applied research, by leveraging recent advances in machine learning literature. Using this indicator, we quantitatively show that cited papers are more likely to be basic research, resonating with earlier results \citep{Narin-linkage-1997, McMillan-biotech-2000}. Yet, we also find that a significant portion of papers cited in patents that are related to FDA-approved drugs are clinical research. These findings appears to be in a sharp contrast with a recent
finding that declares no relationship about whether basic or applied research
are more likely to be cited by patents \citep{Li-applied-2017}. The
inconsistency may be due to the difference in entities analyzed. While we
focused on papers, they focused on grants and made the basic/applied dichotomy
based on grant abstracts. Furthermore, it remains to be seen to what extent one
short grant abstract can represent the actual research performed and how different
the level scores are for papers produced under the same grant.

Throughout the work, we have grouped patents based on NBER categories, which rely on the USPC codes. USPTO, however, scheduled to replace USPC codes with the Cooperative Patent Classification (CPC) schema in 2013 (\url{https://www.uspto.gov/patents-application-process/patent-search/classification-standards-and-development}), raising the question of whether our analysis can be extended to later patents without USPC codes. In Appendix B, we demonstrate that it is still feasible to assign USPC classes and NBER categories to those patents, through their IPC classes.

Some of our analysis may be limited by the quality of bibliographic data of papers. To be specific, we have relied on the first author's affiliation to infer the country and institution type of a paper, due to the unavailability of the affiliation information for other authors. As scientific collaboration has become the dominant mode in knowledge production, future work is needed to collect missing affiliation data and examine how the results may change. Second, we have used the funding information provided in the MEDLINE database to analyze the role of NIH. Although it remains unclear about the completeness of the data, our results nevertheless provide a lower bound on the fraction of cited papers that are funded by NIH. Future studies could also use other data sources such as Web of Science to get the funding information.

Future work is needed to model patent-to-paper knowledge flow among different types of institutions and compare how it is different from patent-to-patent knowledge flow. That the science linkage is dominated by biotechnology and drug patents may suggest a finer level categorization of these patents that goes beyond existing schemes is needed. Future work can base the linkage to science to cluster these patents and compare how the data-driven derived clusters align with traditional schemes.

\appendix

\setcounter{figure}{0}
\makeatletter 
\renewcommand{\thefigure}{A.\@arabic\c@figure}
\makeatother

\setcounter{table}{0}
\makeatletter 
\renewcommand{\thetable}{A.\@arabic\c@table}
\makeatother

\section{Country origin of patents based on first inventor residence}

Table~\ref{tab:top-country-first} presents the number of patents by country based on the residence of first inventor, a standard practice adopted by organizations like WIPO (\url{https://www.wipo.int/ipstats/en/statistics/patents/wipo_pub_931.html}), patent offices such as USPTO (\url{https://www.uspto.gov/web/offices/ac/ido/oeip/taf/appl_yr.htm}) and EPO (\url{https://www.epo.org/about-us/annual-reports-statistics/annual-report/2018/statistics/granted-patents.html\#tab1}), and the literature \citep{Hall-nber-2001, Bacchiocchi-knowledge-2009}.

\begin{table}[h!]
\centering
\caption{Top 20 countries (territories) with most patents.}
\label{tab:top-country-first}
\begin{tabular}{c r r || c r r}
Country & Patents & \% & Country & Patents & \% \\ \hline
US & $605,875$ & 55.7 & TW & $9,849$ & 0.9 \\
JP & $156,491$ & 14.4 & SE & $9,192$ & 0.8 \\
DE & $87,127$ & 8.0 & BE & $7,883$ & 0.7 \\
FR & $35,505$ & 3.3 & IL & $6,825$ & 0.6 \\
GB & $33,274$ & 3.1 & AU & $6,388$ & 0.6 \\
CA & $20,691$ & 1.9 & DT & $5,361$ & 0.5 \\
CH & $17,865$ & 1.6 & JA & $4,847$ & 0.4 \\
KR & $14,901$ & 1.4 & DK & $4,816$ & 0.4 \\
IT & $14,376$ & 1.3 & FI & $4,051$ & 0.4 \\
NL & $10,794$ & 1.0 & AT & $3,476$ & 0.3 \\
\end{tabular}
\end{table}

\section{Assigning USPC classes and NBER categories to patents without USPC codes}

To assign NBER categories to patents without IPC classes, we first establish the mapping from IPC to USPC classes, using the US-to-IPC8 Concordance provided by USPTO. For example, the table that maps USPC subclasses of class 424 to IPC subclass and group can be found at \url{https://www.uspto.gov/web/patents/classification/uspc424/us424toipc8.htm}. We scraped all these tables and created the IPC-to-USPC mapping using fractional counting. As examples, IPC ``A61K 51/00'' is uniquely mapped to USPC 424. IPC ``A61M 36/14'' can be mapped to 22 unique USPC subclasses, 21 of which corresponds to USPC class 424 and the remaining 1 to class 427. Therefore, ``A61M 36/14'' is mapped to USPC 424 with weight $\frac{21}{22}$, and to USPC 427 with weight $\frac{1}{22}$. Table~\ref{tab:ipc} provides the mapping for the 3 exemplar IPC codes.

\begin{table}[h!]
\centering 
\caption{Mapping from IPC code to USPC class, with weight in parentheses.}
\label{tab:ipc}
\begin{tabular}{l l rrrrr}
\hline
IPC code & USPC class and weight \\
\hline
A61K 5100 & 424 (1) \\
A61M 3614 & 424 ($\frac{21}{22}$); 427 ($\frac{1}{22}$) \\
A61K 5104 & 534 (1) \\
\hline
\end{tabular}
\end{table}

For a particular patent without USPC codes, we can then assign USPC classes based on its IPC codes. For example, patent 9,044,520 has 3 IPC codes: ``A61K 5100'', ``A61M 3614'', and ``A61K 5104''. The weight for USPC 424 is $\frac{1}{3} \times 1 + \frac{1}{3} \times 21/22 = 0.652$, similarly the weight for USPC 534 is 0.333, and for USPC 427 is 0.015. Therefore, we assign USPC 424, and accordingly NBER subcategory ``31'' (Drugs) and category ``3'' (Drugs \& Medical) to this patent.

\begin{figure}[ht!]
\centering
\includegraphics[trim=0mm 4mm 0mm 0mm, width=\columnwidth]{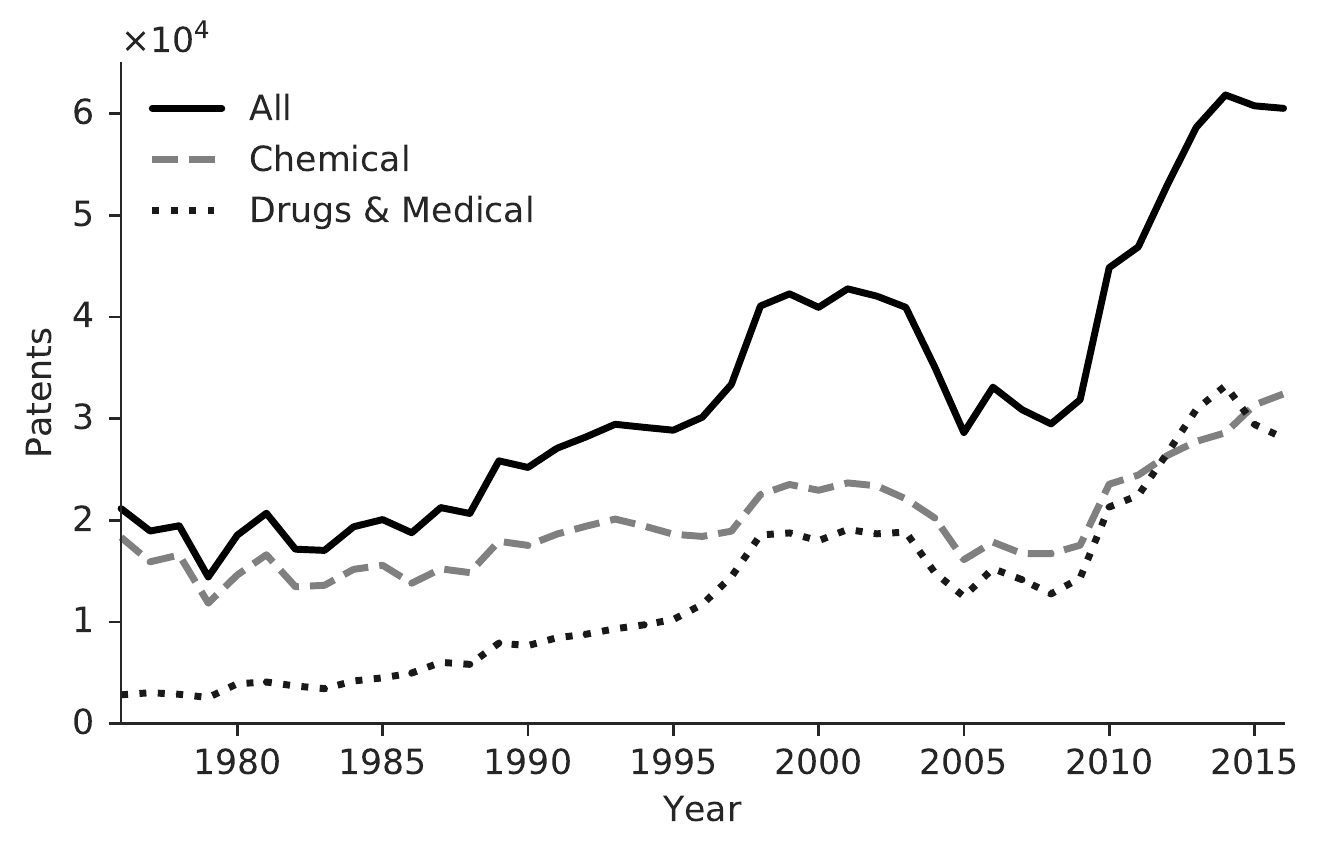}
\caption{The number of patents granted between 1976 and 2016.}
\label{fig:num-pat-2}
\end{figure}

To validate this method, we first note that although USPC was scheduled to be replaced by CPC starting from 2013, patents granted until mid 2015 still have associated USPC classes. We therefore additionally downloaded and parsed bibliographic data for patents granted between 2013 and 2016. We then leverage the fact that patents granted before 2015 have USPC (hence NBER categories) information, which serve as our ground-truth data, and apply the method to patents between 1976 and 2014. We find that for the vast majority (86\%) of these patents, their NBER categories assigned by our method are identical to those based on their USPC. To apply this method in scale, we perform the NBER category assignment for patents granted between mid 2015 and 2016. Fig.~\ref{fig:num-pat-2} displays the number of patents between 1976 and 2016, demonstrating the feasibility of extending our analysis to later patents. 

\end{document}